\documentclass[preprint,showpacs,preprintnumbers,amsmath,amssymb,prb]{revtex4}
\usepackage{graphicx}
\usepackage{dcolumn}
\usepackage{bm}
\usepackage{epsfig}
\begin{document}

\title{Amphiphile Adsorption on Rigid Polyelectrolytes}
\author{Paulo S. Kuhn}
\affiliation{Departamento de F\'{\i}sica, Instituto de F\'{\i}sica e
Matem\'atica,
Universidade Federal de Pelotas,
Caixa Postal 354, 96010-900, Pelotas, RS, Brazil}

\author{Yan Levin}
\affiliation{ Instituto de F\'{\i}sica, Universidade Federal do Rio
Grande do Sul,
Caixa Postal 15051, 91501-970, Porto Alegre, RS, Brazil}
\email{levin@if.ufrgs.br}

\author{Marcia C. Barbosa}
\affiliation{ Instituto de F\'{\i}sica, Universidade Federal do Rio
Grande do Sul,
Caixa Postal 15051, 91501-970, Porto Alegre, RS, Brazil}
\email{marcia.barbosa@ufrgs.br}

\author{Ana Paula Ravazzolo}
\affiliation{Departamento de Patologia Cl\'{\i}nica Veterin\'aria, Instituto
de Veterin\'aria,  Universidade Federal do Rio
Grande do Sul, Porto Alegre, RS, Brazil}

\begin{abstract}

A theory is presented which quantitatively accounts for the cooperative adsorption of cationic surfactants to anionic polyelectrolytes. For high salt concentration we find that the critical adsorption concentration (CAC) is a bilinear function of the polyion monomer and salt concentrations, with the coefficients dependent only on the type of surfactant used. The results presented in the paper might be useful for designing more efficient gene delivery systems. 

\end{abstract}

\maketitle

\section{Introduction}
\label{introduction}

Polyelectrolyte solutions are ubiquitous. Over the years they have attracted attention of Chemists, Physicists, and Biologists. These complex systems have also found many industrial applications ranging from water treatment to superabsorbants. Besides the charged polymers (polyions), a polyelectrolyte solution can contain salt and other molecules which may strongly interact with the polyions resulting in a plethora of interesting behavior. The complexity of polyelectrolyte solutions and the long range of the Coulomb interaction makes study of these systems particularly challenging ~\cite{Levin1,Netz1,Netz2,Low,Vries2,Netz3,Allen,Vries,Khok,Muthu1,Beck,Cruz,Dobrynin,Stevens,Kumar,Muthu2,Pizio,And1,Sain,Piguet,Vries3,Paez}.

In this work we will explore polyelectrolyte solutions containing salt and ionic surfactants.  These systems have attracted particular attention because the polyelectrolyte-surfactant, or more realistically, polyelectrolyte-lipid adsorption can be used to neutralize the DNA charge, facilitating the transfection of 
the DNA across a phospholipid cell membrane. Since the polyions which are of particular interest to us, such as the DNA, are quite rigid (persistence length 50nm) one can model them as rigid cylinders with the monomeric charge uniformly distributed along the main axis.  This provides a significant simplification, since for such rigid molecules the internal degrees of
freedom can be ignored.
 
Association between the cationic surfactants and the anionic polyions is driven by the electrostatic and the  hydrophobic interactions. The hydrophobicity of the surfactant tails is responsible for the {\it cooperative} nature of the surfactant adsorption~\cite{SaYa76,Ku98,Ku99b}. When the concentration of surfactant inside the solution is low, most of the molecules remain free, unassociated. As the concentration of surfactant is increased, a density is reached at which the number of adsorbed  surfactants rises sharply. We will denote this as the critical adsorption concentration (CAC). The goal of the theory is to predict the value of the CAC as a function of the polyelectrolyte and the salt concentrations for different kinds of polyion and surfactant molecules.   

All the information about the thermodynamics of the system is contained in its free energy. The Helmholtz free energy for the polyelectrolyte-surfactant solution can be constructed based on the fundamental ideas of Debye, H\"uckel and Bjerrum\cite{DH1,McQ,Bj,Manning}. The resulting Debye-H\"uckel-Bjerrum theory corrects for the linearization of the
Boltzmann factor by explicitly introducing into the theory the non-linear configurations in the form of clusters composed of
associated anionic and cationic entities.  The concentration of clusters is determined by the law of mass action. It can be shown that in the infinite dilution limit this theory yields the exact Manning limiting laws for rigid polyelectrolyte solutions~\cite{Le96}.

To model the hydrophobic interaction of surfactant tails, we will introduce a hydrophobicity  
parameter $\chi$. The basic theory has been presented elsewhere~\cite{Ku98,Ku99b}, and in this paper we will review it only briefly, referring the interested reader to the original publications.  The goal of the present work is to compare the predictions of the theory with the experimental data on the association of undecyl-, dodecyl, tridecyl-, and tetradeculpyridinium cations with the anionic polyelectrolyte sodium dextran sulfate in an aqueous solutions \cite{Ma84}. We will also derive a simple asymptotic law for the behavior of the CAC at high salt
concentrations, which we expect to be valid for all rigid polyelectrolyte-surfactant systems. Although dextran sulfate is not
nearly as rigid as the DNA for which the original theory was developed, nevertheless, if we restrict ourselves to sufficiently
large salt concentration for which the Debye length is shorter than the polyelectrolyte persistence length, the polyion  flexibility can be neglected. We should also note that the present theory is only valid for sufficiently 
dilute polyelectrolyte solutions prior to a phase separation which can occur following  
addition of cationic surfactants.  The high density phases will have to be described by a different
theory~\cite{Br98}. 

In the next section we will briefly review the thermodynamic model used to perform the calculations.
\begin{figure}
\begin{center}\includegraphics[clip=true,scale=0.4]{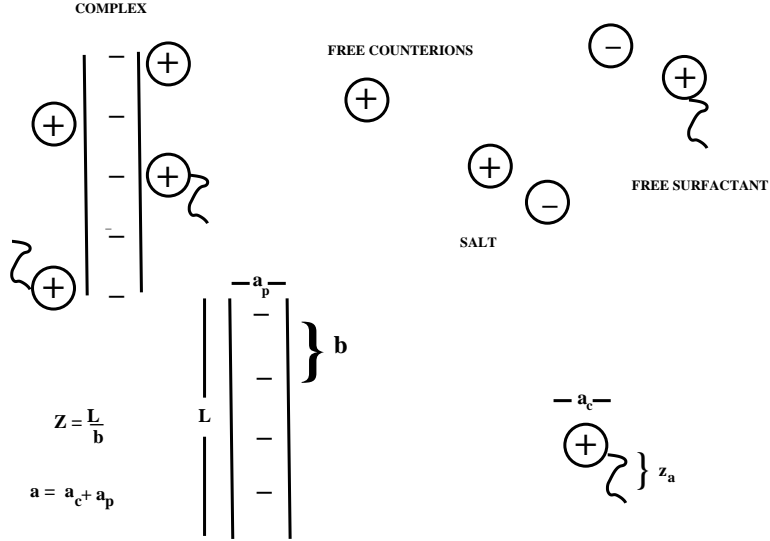}
\end{center}
\caption{The model.}
\label{fig1}
\end{figure}

\section{The Theory}
\label{model}

The polyelectrolyte solution, illustrated in Fig. 1, consists of $N_p$ cylindrical anionic polyions of length $L$ and diameter $a_{p}$, $N_a$ monovalent amphiphiles, and $N_s$ ``molecules'' of salt. In an aqueous solution, the anionic polyions are ionized resulting in $Z$ monomers, each of charge $-q$, distributed uniformly with separation $b=L/Z$ along the main axis.  As a result of ionization,  
$Z$ monovalent counterions are liberated into the solution for each polyion. We will suppose that salt is a strong electrolyte which is
fully dissociated in an aqueous environment resulting in an equal number of positively and negatively charged monovalent ions. Similarly, we will assume that the head group of each surfactant molecule is fully dissociated, producing a negative monovalent coion (charge $-q$) and a polymeric chain with a cationic head group of charge $+q$. For simplicity, all the counterions and coions will be treated as identical, 
independent of the molecules from which they were derived. The ions will be idealized as hard spheres of diameter $a_c$ and charge $\pm q$ located at the center. The surfactant molecules will be modeled as polymers of $z_a$ spherical monomers (each of a diameter $a_c$) 
with the head monomer carrying a charge $+q$. The solvent, water, will be treated as a continuum of dielectric constant $D$. The hydrophobic interaction between the surfactant tails is short ranged and will be characterized by a parameter $\chi$. 

Inside the polyelectrolyte solution, the concentration of polyions is $\rho_p=N_p/ V$, the concentration of monovalent salt is $C_{s}=N_s/V$, and the concentration of monovalent amphiphile (surfactants) is $C_{a}=N_a/V$. The strong electrostatic interactions between the polyions the microions and the cationic surfactants leads to formation of complexes which --- in thermodynamic equilibrium --- will be made of one polyion, $n_c$ associated monovalent counterions, and $n_a$ associated amphiphiles. Particle conservation
requires that 
\begin{eqnarray}
& & \rho_c = (Z - n_c) \, \rho_p + C_s \; , \nonumber \\
& & \rho_{a} = C_a - n_a \, \rho_p \; ,
\label{eq1}
\end{eqnarray}
where $\rho_c$ is the concentration of free monovalent counterions and $\rho_{a}$ is the concentration of free amphiphiles. The concentration of coion $\rho_-$ inside the solution is unaffected by the association,
\begin{eqnarray}
\rho_- = C_s + C_a \; ,
\label{eq2} 
\end{eqnarray}
The total concentration  of free monovalent charges is then
\begin{eqnarray}
\rho_f = \rho_c + \rho_{a} + \rho_- \; .
\label{eq3} 
\end{eqnarray}
The goal of the theory is to determine the number of condensed counterions $n_{c}$ and the condensed surfactants $n_{a}$. In order to achieve this, 
the Helmholtz free energy of the polyelectrolyte solution is constructed and minimized. The details of the calculation can be found elsewhere~\cite{Ku98,Ku99b} here we just give the few main steps. 

The largest contributions to the Helmholtz free energy of a dilute polyelectrolyte solution are electrostatic and entropic,
\begin{equation}
F = F_{el} + F_{ent} \; .
\label{eq4}
\end{equation}
The electrostatic free energy density $f_{el}= F_{el}/V$ is the result of the interaction between the complexes and the free charges inside the solution. It can be obtained using the framework of the Debye-H\"uckel-Bjerrum (DHBj) theory~\cite{LeFi96,Levin1} and has been calculated in our previous work to be~\cite{Ku98,Ku99b}:
\begin{eqnarray}
\beta \, f^{el} = - \frac{\rho_{p} \, Z_c^2 \, (a/L)} {T^{\ast}(\kappa a)^2}
\left[2 \, \ln \, \left[\kappa \, a \, K_1(\kappa \, a) \right] - {\mathcal I}_0(\kappa a) + \frac{(\kappa \, a)^2} {2}\right] \; ,
\label{eq5}
\end{eqnarray}
where $\beta = 1/k_B \, T$, $f = F/N_p$ and
\begin{equation}
{\mathcal I}_0(\kappa a) \equiv \int_0^{\kappa \, a} \, \frac {x \, K_0^2 \, (x)} {K_1^2 \, (x)} \, dx \; .
\label{eq6}
\end{equation}
Here $K_n(x)$ is the modified Bessel function of order $n$, $\kappa$ in $(\kappa \, a)^2 = 4 \, \pi \, \rho_f^*/T^*$ is the inverse of the 
Debye screening length, $\rho_f^* = \rho_f \, a^3$ is the reduced density, and $T^* = D \, k_B \, T \, a/q^2$ is the reduced temperature. 
We have defined $a = (a_c + a_p)/2$ as the effective radius and $Z_c = Z - n_a - n_c$ as the effective charge of the polyion-amphiphile complex. The entropic free energy density is given by \cite{Flory71}
\begin{equation}
\beta f^{ent} = \sum_i \left[\rho_i - \rho _i \, \ln \, \left(\frac{\phi_i} {\zeta_i} \right) \right] \; ,
\label{eq7}
\end{equation}
where $i \in \{p,c,a,-\}$ represents the different species and $\zeta_{i}$ is the internal partition of the specie $i$. For structureless particles, the internal partition function is $\zeta_ - = \zeta_c = 1$. For surfactants, the Flory theory gives $\zeta_{a+} = z_a$. The respective volume fractions $\phi_i$ are:

\begin{eqnarray}
\phi_p & = & \frac{\pi \, \rho_p^{\ast}} {4 \, (a/L)} \, \left(\frac{a_p} {a} \right)^2 \nonumber \\
& & + \frac{Z \, \pi \, \rho_p^{\ast}} {6} \, (z_a \, m_a + m_c) \, \left( \frac{a_c} {a} \right)^3  \, ,\nonumber \\
\phi_c & = & \rho_c^{\ast} \, \frac{\pi} {6} \, \left(\frac{a_c} {a} \right)^3 \, , \nonumber \\
\phi_{a} & = & \frac{z_a \, \pi \, \rho_{a}^{\ast}} {6} \, \left(\frac{a_c} {a}
\right)^3 \, , \nonumber \\
\phi_- & = & \frac{\pi \, \rho_-^{\ast}} {6} \, \left(\frac{a_c} {a} \right)^3 \, .
\label{eq8}
\end{eqnarray}
Here we have introduced the fractions of the polyion monomers associated with the condensed counterions $m_c = n_c/Z$ and with the condensed surfactants $m_a = n_a/Z$. The internal partition function of a complex $\zeta_p$ can be calculated by modeling the polyion as a one dimensional lattice of $Z$ adsorption sites. The number of associated counterions/surfactants at each site can be either zero or one. The problem of calculating the internal partition function of the polyion-amphiphile complex then reduces to finding the free energy of a one dimensional lattice gas of three different states: empty, associated with a counterion, and associated with a surfactant. This free energy can be calculated approximately~\cite{Ku99b} to be

\begin{eqnarray}
- ln \, \zeta_p \left[ m_c,m_a \right] & = & \xi \, K \, \left[ \frac{Z_c} {Z^2}^2 - 1 \right] + \beta \, \chi \left( Z - 1 \right) \, m_a^2 \nonumber \\
& + & Z \, m_c \, \ln \, m_c + Z \, m_a \, \ln \, m_a \nonumber \\
& + & Z \, \left( 1 - m_c - m_a \right) \, \ln \, \left( 1 - m_c - m_a \right) \; ,
\label{eq9}
\end{eqnarray}
where $\xi \equiv \beta \, q^2/D \, a$ is the Manning parameter, $K = Z \, [\psi \, (Z) - \psi \, (1)] - Z + 1 $, and $\psi \, (n)$ 
is the digamma function. The first term of Eq.~(\ref{eq9}) accounts for the electrostatic interaction between the polyion and the condensed counterions and surfactants, the second term is due to the gain in the hydrophobic free energy when two surfactant tails are in a vicinity of each other,  the other terms are  entropic -- resulting from the thermal diffusion of the condensed particles along the polyion chain. The characteristic energy of the interaction between the condensed surfactants is measured by a phenomenological
parameter $\chi$. For each type of surfactant molecule, the value of $\chi$ will be determined by the best fit to the experimental data.   

The equilibrium configuration of the system is found by minimizing the total Helmholtz free energy with respect to the number of associated counterions and surfactants,  
\begin{eqnarray}
\frac{\partial \, F} {\partial \, m_c}  & = & 0  \; , \nonumber \\
\frac{\partial \, F} {\partial \, m_a}  & = & 0  \; .
\label{eq10}
\end{eqnarray}
Solving this system of equations, the number of condensed counterions and surfactants can be determined as a function of the concentration of polyelectrolyte, salt, and surfactant.

\section{Results}
\label{results}

We will define a surfoplex as a complex in which almost all the polyion monomers are neutralized by the cationic amphiphiles. Surfoplexes are formed when the concentration of the amphiphile inside the solution reaches the CAC, $C_a^c$. For practical applications in gene therapy, the CAC is of particular importance since it gives the minimum amount of surfactant needed to neutralize the DNA charge for the transfection across the cell membrane~\cite{Gr06}. Since the cationic surfactants and lipids are quite toxic, in vivo applications require that the amount of amphiphile used in a transfection be as low as possible~\cite{Gr06}.

The CAC depends on the amount of salt inside the solution and on the specifics of the polyelectrolyte and the surfactant to be used. To calculate it, we first  obtain the full adsorption isotherm, from which the surfoplex formation and 
the CAC are identified as the point of the
cooperative binding transition characterized by a sharp rise in the adsorbed fraction of the amphiphile.  We note that in general this can be either a first or a second order phase transition, or simply a sharp crossover.  
\begin{center}
\begin{table}
\begin{tabular}{|c|c|} \hline \hline
$z_a$ & $\chi$ \\ \hline
$12$ & $-5.5$ \\
$13$ & $-6.8$ \\
$14$ & $-7.9$ \\
$15$ & $-9.3$ \\ \hline \hline
\end{tabular}
\caption{\small Relation between the fitted $\chi$ and $z_a$.}
\end{table}
\end{center}

The theory was compared with the experiments of Malovikova et al.\cite{Ma84}. who measured the adsorption isotherms for sodium dextran sulfate for various concentrations of sodium chloride and four different kinds of cationic surfactant: undecyl-, dodecyl-, tridecyl-, and tetradeculpyridinium. The four surfactants studied differ by the length of their hydrocarbon tail ($z_a - 1 = 11, 12, 13$ and $14$) and, consequently, have different values of the hydrophobicity parameter $\chi$. Theoretically~\cite{Ku98}, we expect $\chi$ to be a linear function of $z_a$.
The experiments were performed at $30^oC$. The diameter of sodium dextran sulfate is $a_p = 7 \, \AA\,$ and the distance between the consecutive charged monomers  is $b = 2.5 \, \AA$ ($\xi \approx 2.8$). The binding isotherms obtained experimentally show the formation of surfoplexes when the surfactant concentration reaches the CAC, $C_a^c \, (z_a, C_s, Z, \rho_p)$, which is strongly dependent on the concentration of all the species and on the type of surfactant used. The CAC was found to increase with the salt concentration. The experimental data also showed that the CAC decreased rapidly with the length of the surfactant hydrophobic tail, see Figure 2. 
\begin{figure}
\begin{center}\includegraphics[clip=true,scale=0.6]{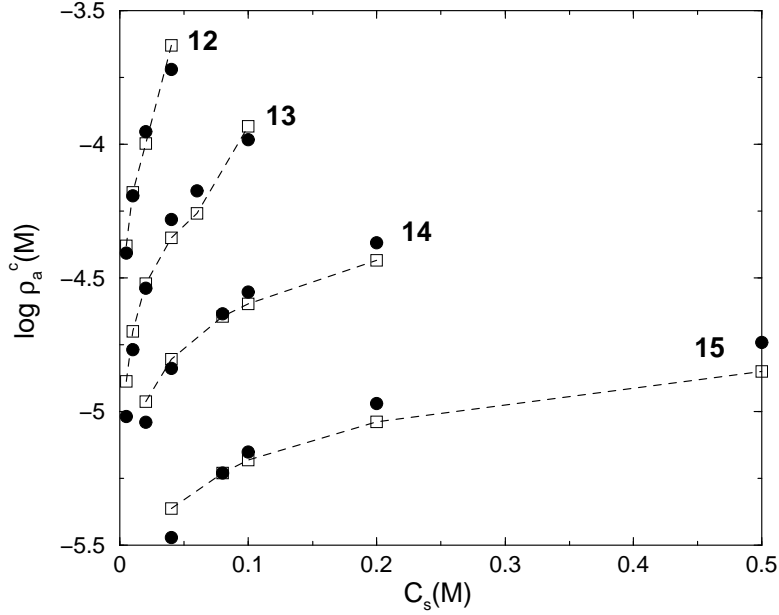}
\end{center}
\caption{Density of {\it free} surfactant at the transition as a function of salt concentration, $\textrm{log}_{10} \, \rho_a^c \, (M) \times C_s \, (M)$ for $z_a = 12, 13, 14, 15$. The filled symbols are the experimental results \cite{Ma84} and the empty symbols are the present theory.  The dashed lines are only a guide to the eyes.}
\label{fig3}
\end{figure}
To compare with the experiment, the theoretical binding isotherms were computed by the procedure described in the previous section. The  number density of polyelectrolyte monomers was taken to be the same as in the reference  \cite{Ma84} . The diameter of the polyion counterions, of salt ions, and of the surfactant monomers were all taken to be  $a_c = 9 \, \AA$. This was done in order to simplify the calculations and to minimize the number of adjustable parameters. The final results were only weakly dependent on the precise value of $a_c$. Each polyion was (arbitrarily) taken to have $Z = 3 \, 000$ charged monomers (experiments do not provide the length of the polyions, but only the number density of the polyion monomers $Z \rho_p$). The hydrophobicity parameter $\chi$ was obtained by performing a fit of the theoretically calculated CACs to the experimental data. For each amphiphile type, a unique value of  $\chi$ was used to obtain the dependence of the CAC on the concentration of salt. Figure 2  shows that the theory agrees quite well with the experimental data of Malovikova et al. for all 4 surfactants.   Table 1 provides the best fit value of the hydrophobicity parameter $\chi$, and Figure 3 shows that $\chi$ is, indeed, a linear function of the surfactant length, as was expected theoretically~\cite{Ku98}. The best fit of $\chi$  yields  
\begin{eqnarray}
\chi = - 1.25 \, z_a + 9.5 \; ,
\label{eq11}
\end{eqnarray}
in units of $k_B T$.
\begin{figure}
\begin{center}\includegraphics[clip=true,scale=0.6]{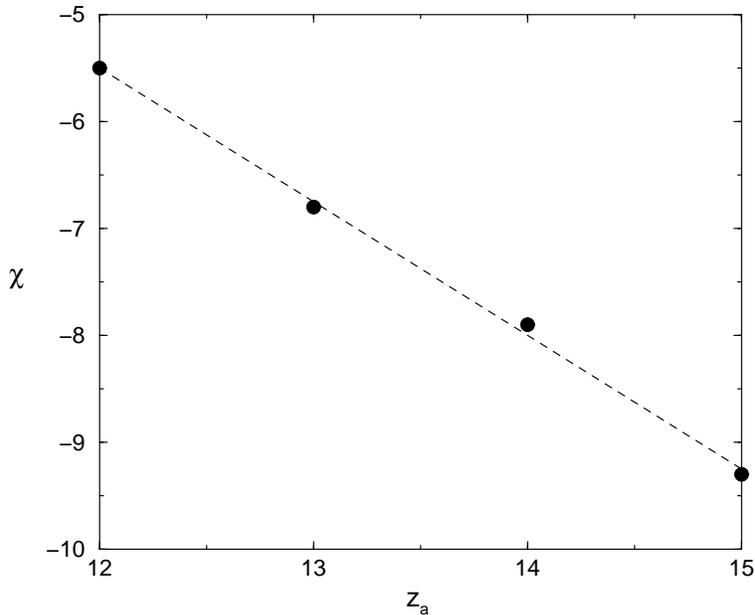}
\end{center}
\caption{The hydrophobicity parameter $\chi$ versus $z_a$ obtained by fitting the experimental data of Malovikova et al.\cite{Ma84}. The dashed line is the linear fit.}
\label{fig4}
\end{figure}

The dependence of the CAC on $C_s$ for various polyion lengths $Z$ at fixed 
monomer concentration $\rho_m = Z \, \rho_p$ is illustrated in Figure 4. 
\begin{figure}
\begin{center}\includegraphics[clip=true,scale=0.6]{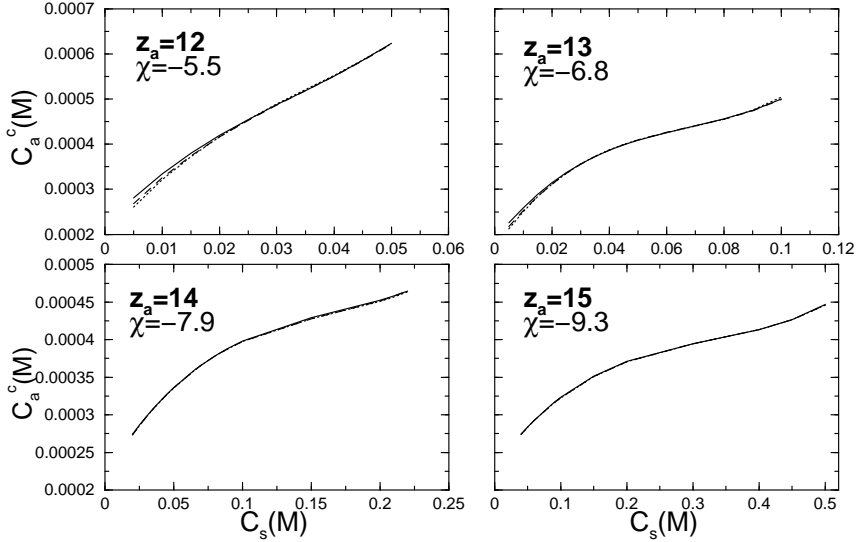}
\end{center}
\caption{Total density of surfactant at the transition (CAC) as function of salt concentration, $C_a^c \, (M) \times C_s \, (M)$ for $\rho_m = Z \, \rho_p = 5 \times 10^{-4} \, M$ kept fixed and $Z = 1 \, 000$ (solid line), $Z = 2 \, 000$ (dashed line), $Z = 3 \, 000$ (dotted line).}
\label{fig5}
\end{figure}
We see that that, as expected, the CAC does not depend individually on $Z$ and $\rho_p$, but only on the product of the two --- the total number density of the polyion monomers $\rho_m$. Furthermore, from Figures 4 and 5 we see that the CAC becomes a {\it linear} function of $C_s$  for { \it sufficiently large} salt concentrations and {\it sufficiently low} monomer density,
\begin{eqnarray}
C_a^c = b_0 \, (z_a, Z \rho_p) + b_1 \, (z_a, Z \rho_p) \, C_s \;.
\label{eq12}
\end{eqnarray}
The amount of salt needed to reach the linear regime depends on the type of surfactant used --- more salt is needed for  surfactants with longer
hydrocarbon tails. We now explore the dependence of the CAC on the concentration of polyion monomers $\rho_m$ inside the linear regime.
\begin{figure}[h]
\begin{center}\includegraphics[clip=true,scale=0.6]{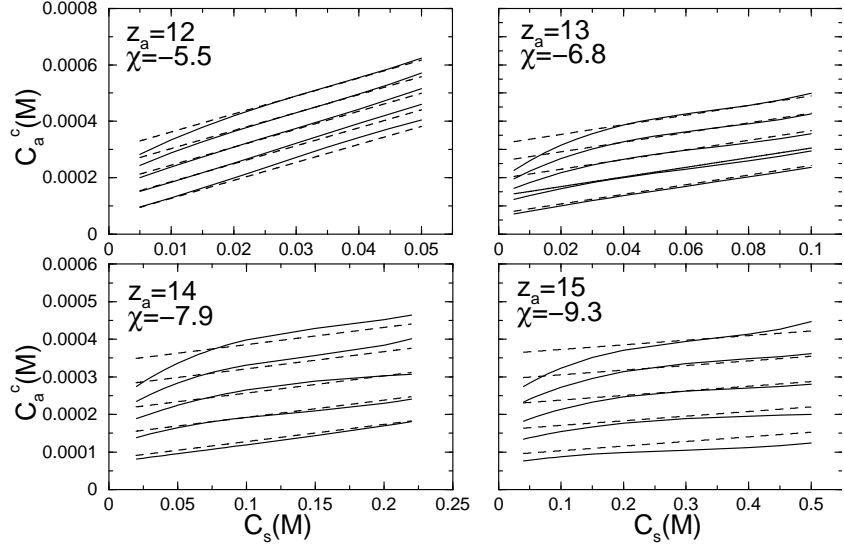}
\end{center}
\caption{The total density of surfactant at the transition as a function of the salt concentration, $C_a^c \, (M) \times C_s \, (M)$ for $Z=1 \, 000$ kept fixed and $\rho_m = 1 \times 10^{-4} \, M$, $2 \times 10^{-4} \, M$, $3 \times 10^{-4} \, M$, $4 \times 10^{-4} \, M$, $5 \times 10^{-4} \, M$ from bottom to top. The dashed lines are the linear adjust, eqs. (\ref{eq13}) and (\ref{eq14}).}
\label{fig6}
\end{figure}
Figure 5 show that while the absolute values of the CAC in the linear regime depend on $\rho_m$, the slope is insensitive to the precise concentration of the polyion monomers. This implies that  $b_1 \, (z_a, Z \rho_p)$ is  independent of $Z \rho_p$ so that for a given polyelectrolyte solution $b_1 \, (z_a, Z \rho_p) = a_1(z_a)$. Furthermore, we note that the CAC lines for uniformly spaced monomer 
concentrations are equidistant, see Figure 5. This implies that the coefficient $ b_0 (z_a, Z \rho_p)$ must be a linear
function of $Z \rho_p$, so that $ b_0 (z_a, Z \rho_p) = c(z_a)+a_0(z_a) Z \rho_p$. The coefficients $b_0$ and $b_1$ obtained using the list square fits of the linear regime shown in the Table 2, clearly support these observations. We thus conclude that for a given polyelectrolyte solution, at sufficiently high salt concentrations and sufficiently low monomer densities, the CAC is a bilinear function of $Z \, \rho_p$ and $C_s$,
\begin{eqnarray}
C_a^c = c(z_a)+a_0(z_a) Z \rho_p + a_1(z_a) \, C_s \;,
\label{eq13}
\end{eqnarray}
with the coefficients dependent only on the length of surfactant used. This is the fundamental result of the present paper. Equation (\ref{eq13}) should be valid for all rigid polyelectrolyte surfactant systems at sufficiently high salt concentrations and sufficiently low monomer densities. For the specific case of the association of undecyl-, dodecyl, tridecyl-, and tetradeculpyridinium surfactants with anionic polyelectrolyte sodium dextran sulfate, the coefficients $c(z_a)$, $a_0(z_a)$, and $a_1(z_a)$ can be obtained by interpolating the data presented in the Tables 2.

We find
\begin{eqnarray}
a_1 \, (z_a) & = & 46 \, 012.89 \, e^{ -1.32 \, z_a } \, ,\nonumber \\
a_0 \, (z_a) & = & 0.25 + 0.028 \, z_a \, , \nonumber \\
c \, (z_a) & = & -7.47 \times 10^{-5} + 6.57 \times 10^{-6} \, z_a \, .
\label{eq14}
\end{eqnarray}

\begin{center}
\begin{table}
\begin{tabular}{|c|c|c|c|} \hline \hline
$\rho_m \, (\times 10^{-4} \, M)$ & $z_a$ & $b_0$ & $b_1$ \\ \hline
$1$ & $12$ & $0.000074$ & $0.0067$ \\
& $13$ & $0.000069$ & $0.0017$ \\
& $14$ & $0.000067$ & $0.00052$ \\
& $15$ & $0.000073$ & $0.0001$ \\ \hline
$2$ & $12$ & $0.00011$ & $0.0069$ \\
& $13$ & $0.00013$ & $0.0016$ \\
& $14$ & $0.00014$ & $0.00044$ \\
& $15$ & $0.00017$ & $0.000073$ \\ \hline
$3$ & $12$ & $0.00016$ & $0.0071$ \\
& $13$ & $0.00021$ & $0.0014$ \\
& $14$ & $0.00026$ & $0.00023$ \\
& $15$ & $0.00025$ & $0.000062$ \\ \hline
$4$ & $12$ & $0.00022$ & $0.007$ \\
& $13$ & $0.00028$ & $0.0014$ \\
& $14$ & $0.00027$ & $0.00058$ \\ 
& $15$ & $0.00028$ & $0.00017$ \\\hline
$5$ & $12$ & $0.00029$ & $0.0067$ \\
& $13$ & $0.00033$ & $0.0016$ \\
& $14$ & $0.00036$ & $0.00048$ \\
& $15$ & $0.00032$ & $0.00024$ \\
\hline \hline
\end{tabular}
\caption{\small Values of $b_0$, $b_1$ in $C_a^c = b_0 + b_1 \, C_s$ with $Z = 1 \, 000$.}
\end{table}
\end{center}

The results for the CAC obtained with the eqs.(\ref{eq13}) and (\ref{eq14}) are shown as the dashed lines in the Figure 5.

\section{The Conclusions}
\label{conclusions}

We have presented a theory which quantitatively accounts for the cooperative adsorption of charged surfactants to rigid polyelectrolytes. The theory is based on the fundamental ideas of Debye, H\"uckel and Bjerrum~\cite{Levin1}. The only free parameter is the magnitude of the hydrophobic interaction between the condensed surfactants $\chi$.  In agreement with the theoretical expectations, this parameter is found to be a linear function of the surfactant length $z_a$.  For high salt concentrations and sufficiently low monomer densities, we find that the CAC is a bilinear function of the monomer $Z \rho_p$ and salt $C_s$ concentrations, with the coefficients dependent only on the length of surfactants used. For the specific case of the association of the undecyl-, dodecyl, tridecyl-, and tetradeculpyridinium surfactants with the anionic polyelectrolyte sodium dextran sulfate, the coefficients of the bilinear form are calculated  explicitly and presented in eqs. (\ref{eq13}) and (\ref{eq14}).  

We note that the present theory does not take into account 
the non-electrostatic specific interactions between the polyions
and surfactants.  In view of a good agreement between the theory and the experiment, however,
we conclude that such interactions are not very
significant, at least for the present polyelectrolyte surfactant system. 
It would be of great interest to test the the predictions of the theory, 
in particular the bilinear dependence of the CAC on the salt and the monomer 
concentrations, on other polyelectrolyte-surfactant systems.

\end{document}